\documentclass[twocolumn,aps,pra]{revtex4}
\usepackage[latin9]{inputenc}
\usepackage{amsmath}
\usepackage{amssymb}
\usepackage{graphicx}

\makeatletter
\@ifundefined{textcolor}{}
{%
 \definecolor{BLACK}{gray}{0}
 \definecolor{WHITE}{gray}{1}
 \definecolor{RED}{rgb}{1,0,0}
 \definecolor{GREEN}{rgb}{0,1,0}
 \definecolor{BLUE}{rgb}{0,0,1}
 \definecolor{CYAN}{cmyk}{1,0,0,0}
 \definecolor{MAGENTA}{cmyk}{0,1,0,0}
 \definecolor{YELLOW}{cmyk}{0,0,1,0}
}

\usepackage{bm}\usepackage{amsthm}\usepackage{epstopdf}\usepackage{txfonts}

\newcommand{\prlsection}[1]{{\em {#1}:--~}}

\DeclareRobustCommand{\openzero}{\leavevmode\hbox{0\kern-.55em0}}
\mathchardef\minus="002D

\makeatother

\begin{document}

\title{Coherent quantum dynamics in \mbox{steady-state} manifolds of strongly
dissipative systems}

\author{Paolo Zanardi}

\author{Lorenzo Campos Venuti}
\affiliation{Department of Physics and Astronomy, and Center for Quantum Information
Science \& Technology, University of Southern California, Los Angeles,
CA 90089-0484}
\begin{abstract}
It has been recently realized that dissipative processes can be harnessed
and exploited to the end of coherent quantum control and information
processing. In this spirit we consider strongly dissipative quantum
systems admitting a non-trivial manifold of steady states. We show
how one can enact adiabatic coherent unitary manipulations e.g., quantum
logical gates, inside this \mbox{steady-state} manifold by adding
a weak, time-rescaled, Hamiltonian term into the system's Liouvillian.
The effective long-time dynamics is governed by a projected Hamiltonian which results from the interplay between the weak
unitary control and the fast relaxation process. The leakage outside
the \mbox{steady-state} manifold entailed by the Hamiltonian term
is suppressed by an environment-induced symmetrization of the dynamics.
We present applications to quantum-computation in decoherence-free
subspaces and noiseless subsystems and numerical analysis of non-adiabatic
errors. 
\end{abstract}
\maketitle
\prlsection{Introduction} Weak coupling to the environmental degrees
of freedom is often regarded as one of the essential prerequisites
for realizing quantum information processing. In fact decoherence
and dissipation generally spoil the unitary character of the quantum
dynamics and induce errors into the computational process. In order
to overcome such an obstacle a variety of techniques have been devised
including quantum error correction \cite{QEC}, decoherence-free subspaces
(DFSs) \cite{DFS,DFS-exp} and noiseless subsystems (NSs) \cite{NS,NS-exp}. However, it
has been recently realized that dissipation and decoherence may even
play a positive role to the aim of coherent quantum manipulations.
Indeed, it has been shown that, properly engineered, dissipative dynamics
can in principle be tailored to enact quantum information primitives
such as quantum state preparation \cite{kastoryano2011dissipative},
quantum simulation \cite{barreiro2011open,gauge} and computation
\cite{verstraete2009quantum}.

In this Letter we investigate the regime where the coupling of the
system to the environment is very strong and the open system dynamics
admits a non-trivial steady state manifold (SSM). We will show how,
in the long time limit, unitary manipulations e.g., quantum gates,
inside the SSM can be enacted by adding a time-rescaled Hamiltonian
acting on the system only. This coherent dynamics is governed by a
sort of projected Hamiltonian which results from the non-trivial interplay
between the weak unitary control term and the strong dissipative process.
The latter effectively renormalizes the former by continuously projecting
the system onto the \mbox{steady-state} manifold and adiabatically
decoupling the non steady states. 
Several of the results of this Letter can be regarded as a rigorous formulation and significant extension
of ideas first explored in \cite{angelo} and \cite{ogy}.
We would also like to point out
the relation with techniques relying on some type of
quantum Zeno dynamics \cite{dan-zeno,exp-zeno,gauge,cave}.
The latter can be in fact regarded as a special case of our general result (\ref{projection-th}).

This Letter  is organized as follows: we first set the stage of our analysis
and describe the main theoretical ideas and results. We then discuss
in detail, aided by numerical simulations, a few different models
demonstrating dissipation-assisted computation over SSMs comprising
decoherence-free subspaces and noiseless-subsystems. For the reader's
convenience we have collected background technical material and all
the mathematical proofs in \cite{SM}.

\prlsection{Evolution of steady state manifolds \label{stage} }
In the following ${\cal H,\,(\mathrm{dim}({\cal H})<\infty)}$ will
denote the Hilbert space of the system and ${\mathrm{L}}({\cal H})$
the algebra of linear operators over it. A time-independent Liouvillian
super-operator ${\cal L}_{0}$ acting on L$({\cal H})$ is given.
The SSM of ${\cal L}_{0},$ comprises all the quantum states $\rho$
contained in the kernel ${\mathrm{Ker}}\,{\cal L}_{0}:=\{X\,/\,{\cal L}_0(X)=0\} $ of ${\cal L}_0.$ 
We will denote by ${\cal P}_{0}$ (${\cal Q}_{0}:=1-{\cal P}_{0}$)
the spectral projection over Ker$\,{\cal L}_{0}$ (the complementary
subspace of Ker${\cal L}_{0}.$). One has that ${\cal P}_{0}^{2}={\cal P}_{0}$
and ${\cal P}_{0}\,{\cal L}_{0}={\cal L}_{0}\,{\cal P}_{0}=0,$ notice
also that ${\cal P}_{0}$ may not be hermitian. 
The Liouvillian ${\cal L}_{0}$ is also assumed to be such that: \textbf{{a)}}
the equation ${\cal E}_{t}^{(0)}:=e^{t{\cal L}_{0}},\,(t\ge0)$ defines
a semi-group of trace-preserving positive maps with $\|{\cal E}_{t}^{(0)}\|\le1$
\cite{norm}; \textbf{{b)}} The non zero eigenvalues $\lambda_{h},\,(h>0)$
of ${\cal L}_{0}$ have negative real parts i.e., the SSM is attractive.
In this case ${\cal P}_{0}=\lim_{t\to\infty}{\cal E}_{t}^{(0)}.$

On top of the  process described by ${\cal L}_{0}$
we now add a control Hamiltonian term ${\cal K}:=-i[K,\,\bullet]$
where $K=K^{\dagger}=T^{-1}\tilde{K}.$ The time $T$ is a scaling
parameter that, in the spirit of the adiabatic theorem, will be eventually
sent to infinity. If $\|\tilde{{ K}}\|=O(1)$ then $\|{\cal K}\|\le2\|K\|=O(1/T).$ 
The basic dynamical equation we are going to study is the following
\begin{equation}
\frac{d\rho(t)}{dt}=({\cal L}_{0}+{\cal K})\rho(t)=:{\cal L}\rho(t).\label{basic-eq}
\end{equation}

 Notice that even if we are {\em{not}} assuming that ${\cal L}_0$ is of the Lindblad type \cite{Lindblad-paper} i.e., the ${\cal E}_t:=e^{t{\cal L}}$ being completely positive (CP) maps, our basic equation (\ref{basic-eq}) is time-local and in this sense Markovian. The system is also strongly dissipative in the sense that, for large $T,$  the dominant process is the  one ruled by ${\cal L}_0$. 
If the system is initialized in one of its steady states, on general
physical grounds one expects the system, for small $1/T,$ to stay
within the SSM with high probability. However, for ${\cal L}_{0}$
with a multi-dimensional SSM a non-trivial internal dynamics may unfold.

In order to gain physical insight on this phenomenon we would like
first to provide a simple argument based on time-dependent perturbation
theory.  Eq.~(\ref{basic-eq}) immediately leads to  $\dot{{\cal E}}_{t}=({\cal L}_{0}+{\cal K}){\cal E}_{t}$ for the evolution semi-group.
We can formally solve this equation by ${\cal E}_{t}=e^{t{\cal L}_{0}}(1+\int_{0}^{t}d\tau\, e^{-\tau{\cal L}_{0}}{\cal K}\,{\cal E}_{\tau})$
from which, by iteration, it follows the standard Dyson expansion 
with respect the perturbation ${\cal K}.$
Considering terms up the first order applied to ${\cal P}_0$ and inserting the spectral resolution ${\cal P}_0+{\cal Q}_0=\mathbf{1}$
one obtains 
${\cal P}_{0}+t\,{\cal P}_{0}{\cal K}{\cal P}_{0}+(e^{t{\cal L}_{0}}-1){\cal S}{\cal K}{\cal P}_{0}$
where ${\cal S}:=-\int_{0}^{\infty}dt \,e^{t{\cal L}_{0}}{\cal Q}_{0}$
is a pseudo-inverse of ${\cal L}_{0}$ i.e., ${\cal L}_{0}{\cal S}={\cal S}{\cal L}_{0}={\cal Q}_{0}.$
The norm of the third term is upper bounded by $O(\|{\cal K}\| \|{\cal S}\|)$
{uniformly} in $t\in[0,\,\infty).$ 
It follows that scaling ${\cal K}$ by $T^{-1},$ over a total evolution
time $t=T$ the first and second term above are $O(1),$ while the
third one --the only one involving transitions outside the steady
state manifold -- is $O(\|{\cal S}\|/T).$ This demonstrates that, at this order
of the Dyson expansion, the dynamics is ruled by an effective generator
${\cal P}_{0}\,{\cal K}\,{\cal P}_{0}$ whose emergence is basically
due to a Fermi Golden Rule mechanism.
Moreover, by looking at the structure of the Liouvillian pseudo-inverse
${\cal S}$ \cite{SM}, we see that $\|{\cal S}\|=O(\tau_R)$
where $\tau_R^{-1}:=\min_{h>0}|\Re\,\lambda_{h}|,$
and  the $\lambda_{h}$'s are the non-vanishing eigenvalues of ${\cal L}_{0}$
\cite{SM}. The meaning of this quantity is
that the time-scale $\tau_R$ sets a lower bound to the relaxation
time of the irreversible process described by ${\cal L}_{0}.$ 
Since $\|{\tilde{K}}\|=O(1)$ if no nilpotent blocks are present in the spectral resolution of ${\cal L}_0$ \cite{kato}  the leakage {outside the SSM} becomes negligible when 
\begin{equation}
T\gg \tau_R
\label{op-time}
\end{equation}
namely  when the time-scale $T$ is much longer  that the relaxation time $\tau_R$ i.e., dissipation is much faster than the coherent part of the dynamics.
System specific examples of (\ref{op-time}) will be given later when concrete applications are discussed.

Now we present our main technical result on the projected dynamics over SSMs (see \cite{SM} for the proof's details):
\begin{equation}
\|{\cal E}_{T}{\cal P}_{0}-e^{\tilde{{\cal K}}_{eff}}{\cal P}_{0}\|=O(1/{T})\label{projection-th}
\end{equation}
 where  $\tilde{{\cal K}}_{eff}:={\cal P}_{0}\,\tilde{{\cal K}}\,{\cal P}_{0}$ and ${\cal E}_{T}$ denotes the evolution over $[0,\, T]$ generated
by ${\cal L}_{0}+T^{-1}\tilde{{\cal K}}.$ 
 It should be stressed
that (\ref{projection-th}) is based just on degenerate perturbation
theory for general linear operators \cite{kato}. In particular, it
does {\em not} rely on the assumption that ${\cal L}_{0}$ can
be cast in Lindblad form \cite{Lindblad-paper} or on the SSM structure
described in \cite{baum}. 
An immediate corollary of (\ref{projection-th}) is that $\|{\cal Q}_{0}{\cal E}_{T}{\cal P}_{0}\|=O(1/T)$,
namely the probability of leaking outside of the SSM, induced by the
unitary term ${\cal K},$ for large $T,$ is smaller than $c\, T^{-1}.$
The constant $c$ controls the strength of the deviations from the
ideal adiabatic behavior at finite $T$ {[}the rhs of (\ref{projection-th}){]}
and it can be related to the spectral structure of ${\cal S}$. Roughly
speaking, one expects $c,$ and therefore violations of adiabaticity,
to increase when the dissipative gap $\tau_R^{-1}$ decreases. However,
a subtler interplay between the gap with the matrix elements of ${\cal Q}_{0}\,{\cal K}\,{\cal P}_{0}$
may play an important role here as well in the information-geometry
of SSM \cite{Banchi}.

Let us now turn to the structure of the effective generator $\tilde{{\cal K}}_{eff}.$
Of course it crucially depends on the projection ${\cal P}_{0}$ that
in turn depends on the nature of ${\cal L}_{0}.$ Here below we discuss
two (non mutually exclusive) cases. Their physical relevance relies on the importance, both theoretical and experimental,
of the concepts of decoherence-free subspaces \cite{DFS} and noiseless-subsystems \cite{NS} in quantum information.

\textbf{{i)}} 
The most general dissipative generator ${\cal L}_{0}$ of a Markovian
quantum dynamical semi-group ${\cal E}_{t}:=e^{t\,{\cal L}_{0}}$
can be written as ${\cal L}_{0}(\rho)=\Phi(\rho)-\frac{1}{2}\{\Phi^{*}(\openone),\,\rho\}$
where $\Phi$ is a CP map $\Phi^{*}$ is the dual map i.e., $\Phi(X)=\sum_{i}A_{i}\rho A_{i}^{\dagger}\Rightarrow\Phi^{*}(X)=\sum_{i}A_{i}^{\dagger}\rho A_{i}$
\cite{Lindblad-paper}. We now assume that $\Phi$ is trace-preserving
($\Phi^{*}(\openone)=\openone$) {\em and} unital ($\Phi(\openone)=\openone$).
Under these assumptions
whence Ker$\,{\cal L}_{0}$ coincides with the set of fixed points
of $\Phi.$ The latter is known to be the commutant ${\cal A}^{\prime}$
\cite{commutant}
of the {\em{interaction algebra}} ${\cal A}$ generated by the Kraus operators $A_{i}$ and
their conjugates \cite{kribs}. From \cite{commutant} it follows
that the SSM of ${\cal L}_{0}$ is $\sum_{J}n_{J}^{2}$-dimensional
and is given by the convex hull of states of the form $\omega_{J}\otimes\openone_{d_{J}}/d_{J}$
where $\omega_{J}$ is a state over the noiseless-subsystem factor
${\mathbf{C}}^{n_{J}}.$ If, for some $J,$ $d_{J}=1,$ one has that
the corresponding ${\mathbf{C}}^{n_{J}}$ is a DFS and the SSM contains
pure states. Conversely, if $d_{J}>1,\,(\forall J)$ then no pure
states are in the SSM. A characterization of the algebraic structure
of SSMs for ${\cal L}_{0}$'s of the Lindblad form \cite{Lindblad-paper}
is provided in \cite{baum}.

Now ${\cal P}_{0}$
is the projection onto the commutant algebra \cite{commutant} and 
 one can check that $\tilde{{\cal K}}_{eff}|_{{\mathrm{Ker}}\,{\cal L}_{0}}=-i[\tilde{K}_{eff},\,\bullet]$
where $\tilde{K}_{eff}:={\cal P}_{0}(\tilde{K})$ \cite{proj-comm}.
By definition $[\tilde{K}_{eff},U]=0$ for all the unitaries in ${\cal A},$ namely the effective dynamics admits as a symmetry group
the full-unitary group of the interaction algebra $\cal A$.
This means that the renormalization process $\tilde{K}\mapsto\tilde{K}_{eff}\in{\cal A}^{\prime}$
amounts to an environment-induced {\em symmetrization} of the dynamics
\cite{symm}. From \cite{commutant}  it also follows that $\tilde{K}_{eff}$
has a non-trivial action just on the noiseless-subsystems of ${\cal A};$ 
the symmetrization process dynamically decouples the system
from the noise process driven by operators in $\cal A$ \cite{symm,ogy}.

\textbf{{ii)}} Suppose there exists a subspace ${\cal C}\subset{\cal H}$
such that ${\mathrm{Ker}}\,{\cal L}_{0}\supset {\mathrm{L}}({\cal C}):=\mathrm{span}\{|\phi_{i}\rangle\langle\phi_{j}|\,/\,|\phi_{i}\rangle\in{\cal C}\}.$
In particular $|\psi\rangle\in{\cal C}\Rightarrow{\cal L}_{0}(|\psi\rangle\langle\psi|)=0$
i.e., ${\cal C}$ is a DFS \cite{DFS} for the unperturbed ${\cal L}_{0}.$
If also ${\cal P}_{0}(|\phi\rangle\langle\phi^{\perp}|)={\cal P}_{0}(|\phi^{\perp}\rangle\langle\phi|)=0$
hold for all $|\phi\rangle\in{\cal C}$ and $|\phi^{\perp}\rangle\in{\cal C}^{\perp},$
a simple calculations shows that ${\cal P}_{0}\tilde{{\cal K}}{\cal P}_{0}|_{{\mathrm{L}}({\cal C})}=-i[\Pi\tilde{K}\Pi,\,\bullet],$
where $\Pi$ is the orthogonal projection over ${\cal C}$ \cite{simple-calcu}.

Remarkably, in all cases \textbf{{i)}}--\textbf{{ii)}} above
we see that the induced SSM dynamics $e^{\tilde{K}_{eff}}$ is {\em
unitary} and governed by a dissipation-projected Hamiltonian.
Qualitatively: this coherent dynamics results from the interplay between
the weak (slow) Hamiltonian $K=T^{-1}\tilde{K}$ and the strong (fast)
dissipative term ${\cal L}_{0}.$ The former induces transitions out
of the SSM while the latter projects the system back into it on much
faster time-scale. As a result non \mbox{steady-state} of the Liouvillian
are adiabatically decoupled from the dynamics up to contributions
$O(1/T)$. We would like now to make a few important remarks. 

\textbf{{1)}} By defining $\tilde{\rho}(t):={\cal U}_{t}^{\dagger}(\rho(t))$
Eq.~(\ref{basic-eq}) gives rise to a dynamical equation  of the
form $d\tilde{\rho}(t)/dt={\cal L}_{t}(\tilde{\rho}(t))$ where ${\cal L}_{t}:={\cal U}_{t}^{\dagger}\circ{\cal L}_{0}\circ{\cal U}_{t}$
and ${\cal U}_{t}(X):=e^{-itK}Xe^{itK}.$ Namely in this rotated frame
$\tilde{\rho}(t)$ evolves in a time-dependent bath described by ${\cal L}_{t}$.
This establishes a connection of the present approach to the one with
time-dependent baths in \cite{angelo} and \cite{ogy}. Smallness
of $K$ in the picture (\ref{basic-eq}) translates into slowness
of the bath time-dependence in the rotated frame.

\textbf{{2)}} The environment-induced renormalization $\tilde{K}\mapsto\tilde{K}_{eff}={\cal P}_{0}K{\cal P}_{0}$
is {\em not} an algebra homomorphism; this implies that the algebraic
structure of a a set of projected Hamiltonians may differ radically
from the algebraic structure of the original (unprojected) ones. In
particular commuting (non-commuting) $\tilde{K}$'s may be mapped
onto non-commuting (commuting) $\tilde{K}_{eff},$ this implying a
potential increase (decrease) of their ability to enact quantum control
\cite{ogy,cave}. Notice that also the Hamiltonian locality structure
may be affected by the projection e.g., a $1$-local $K$ may give
rise to a $3$-local $K_{eff}.$ The dissipative technique here discussed
might then be exploited to effectively generate non-local interactions
out of simpler ones in a fashion similar to perturbative gadgets \cite{gadgets}
(see also \cite{gauge}).

\textbf{{3)}} Any extra term ${\cal V}$, either Hamiltonian or
dissipative, in the Liouvillian such that ${\cal P}_{0}\,{\cal V}\,{\cal P}_{0}=0$
will not contribute to the effective dynamics (\ref{projection-th})
in the limit in which ${\cal L}_{0}$ dominates. For example in the
case \textbf{{ii)}} discussed in the above the projected dynamics
does not change by perturbing $K$ with any extra Hamiltonian term
$K^{\prime}$ such that $\|K^{\prime}\|=O(1/T)$ and ${\cal P}_{0}(K^{\prime})=\sum_{J}{\mathrm{Tr}}_{d_{J}}\left(\Pi_{J}\, K^{\prime}\,\Pi_{J}\right)\otimes\openone_{d_{J}}/d_{J}=0$
{[}here $\Pi_{J}$ is the projector $\openone_{n_{J}}\otimes\openone_{d_{J}}$
of the $J-$th summand in \cite{commutant} {]}. 
The projected dynamics has a degree of resilience against perturbations
that are eliminated by the environment-induced symmetrization.

\textbf{{4)}} If the interaction algebra ${\cal A}$ in {\bf{ii)}} is an Abelian then from \cite{commutant} one finds ${\cal P}_{0}(K)=\sum_{J} \Pi_{J}\, K\,\Pi_{J}.$ This shows that  quantum Zeno dynamics and the associated control and computation techniques of Refs.~\cite{dan-zeno, exp-zeno, cave} can be regarded as a {\em{special}} case of the projection phenomenon described by Eq.~(\ref{projection-th}).

\textbf{{5)}} When ${\cal P}_{0}{\cal K}{\cal P}_{0}=0$ the Dyson series for ${\cal E}_t$ shows that, 
${\cal E}_t {\cal P}_0= 1-t \,{\cal P}_0 {\cal K} {\cal S} {\cal K} {\cal P}_0 +O(\|{\cal K}\|\|{\cal S}\|).$
This means that the dynamics inside the SSM is now ruled by the second-order effective generator
${\cal L}_{eff}:=-{\cal P}_{0}{\cal K}{\cal S}{\cal K}{\cal P}_{0},$ (up to errors $O(\tau_R \|K\|)$).
This dynamics is in general non unitary and its effective relaxation time can be roughly estimated by $\tau^{eff}_R=O(\|{\cal L}_{eff}\|^{-1})
= O( \tau_R^{-1} \|{\cal K}\|^{-2})=\tau_R O((\tau_R\|{\cal K}\|)^{-2})\gg \tau_R.$
Notice the counterintuitive fact that the stronger the dissipation outside the SSM the {\em{weaker}} the effective one inside \cite{angelo}.
\begin{figure}
\noindent \begin{centering}
\includegraphics[width=4cm,height=3cm]{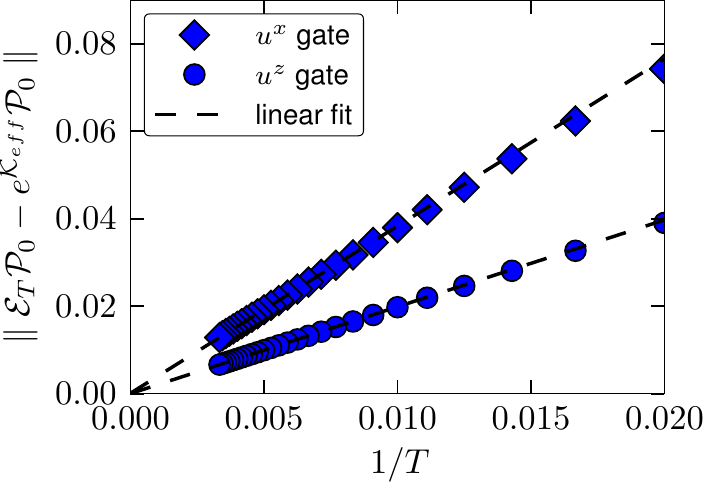} \includegraphics[width=4cm,height=3cm]{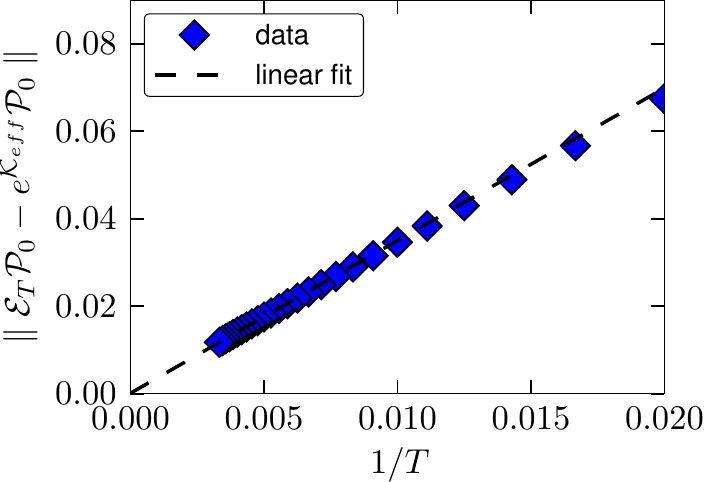} 
\par\end{centering}

\protect\caption{(Color online) Distance of the effective evolution from the exact
one as a function of $1/T.$ Left panel: DFS example with $\mathcal{L}_{0}$
given by Eq.~(\ref{L0-four}). We used $\gamma_{\alpha}=1$ and $\vartheta=1$
(see text for details). Right panel: example for noiseless subsystem
with parameters $\phi_{j}=\vartheta=1$ (see text). The norm used
is the maximum singular value of the maps realized as matrices over
${\cal H}^{\otimes\,2}$. The linear fits are obtained using the four
most significant points. \label{fig:J_zero}}
\end{figure}

\prlsection{Unitaries over a DFS. } We show here how to perform
coherent manipulations on a logical qubit built upon the SSM of four
qubits which comprises a DFS \cite{DFS}. Consider the following unperturbed
Liouvillian 
\begin{equation}
\mathcal{L}_{0}\left(\rho\right)=\sum_{\alpha=x,y,z}\gamma_{\alpha}\left(S^{\alpha}\rho S^{\alpha\dagger}-\frac{1}{2}\left\{ S^{\alpha\dagger}S^{\alpha},\rho\right\} \right)\label{L0-four}
\end{equation}
 where $S^{\alpha}=\sum_{j=1}^{N}S_{j}^{\alpha}$ are collective spin
operators and $\gamma_{\alpha}$ decoherence rates. 
The interaction algebra $\cal A$ generated by the $S^{\alpha}$'s 
is the algebra of permutation invariant operators \cite{NS}.
Therefore, from {\bf{i)}}, it 
follows that Ker${\cal L}_{0}$ has the structure \cite{commutant}
where $J$ is now a total angular momentum label, $d_{J}=2J+1$ and
the $n_{J}$'s are the dimensions of irreps of the permutation group
${\cal S}_{N}$ \cite{perm-dim}. For $N=4$ the one-dimensional $J=0$
representation shows with multiplicity two. If we denote by ${\cal C}$
this two-dimensional  subspace 
the conditions in \textbf{{ii)}} are met. 

Let us denote with $\Pi$
the projector onto ${\cal C}$.
It is known that one can construct universal set of gates in similar
DFSs (see e.g.~\cite{daniel_singlet}) when the dynamics is entirely
contained in the DFS. Here we show that coherent manipulation is possible
also when the dynamics leaks out of the DFS. Consider for example
the following Hamiltonian perturbations 
$H^{x}=\frac{3}{2}\left(\sigma_{1}^{z}\sigma_{2}^{z}+\sigma_{2}^{z}\sigma_{3}^{z}\right)+\openone$
and $H^{z}=-\frac{\sqrt{3}}{2}\left(\sigma_{1}^{z}\sigma_{2}^{z}-\sigma_{2}^{z}\sigma_{3}^{z}\right)+\sigma_{1}^{z}.$
One can check that in the logical space ${\cal C}$, such Hamiltonians
reduce to elementary Pauli operations, i.e.~$\Pi H^{\alpha}\Pi=\sigma^{\alpha}$.
We now build the perturbed Liouvillians $\mathcal{L}^{\alpha}=\mathcal{L}_{0}-i\vartheta/T\left[H^{\alpha},\bullet\right]$,
$\alpha=x,z$, let us also denote $\tilde{\mathcal{K}}_{eff}^{\alpha}=-i\vartheta\mathcal{P}_{0}\left[H^{\alpha},\bullet\right]\mathcal{P}_{0}$
with $\vartheta$ free parameter. In Fig.~\ref{fig:J_zero} left panel we show
a numerical experiment confirming our general theorem Eq.~(\ref{projection-th})
for such $\mathcal{L}^{\alpha}$. In the logical qubit space, the
effective evolution $e^{\tilde{{\cal K}}_{eff}^{\alpha}}$ is a unitary
evolution $e^{\tilde{{\cal K}}_{eff}^{\alpha}}(X)\simeq u^{\alpha}Xu^{\alpha\dagger}$
with $u^{\alpha}=\exp(-i\vartheta\sigma^{\alpha})$, and one can easily
generate any unitary in $SU(2)$ by concatenating such gates. Moreover,
the bound in Eq.~(\ref{projection-th}) implies that, for any vectors
$|i\rangle,\,|j\rangle$ in the logical space ${\cal C}$, $\parallel(\mathcal{E}_{T}-e^{\tilde{\mathcal{K}}_{eff}})(|i\rangle\langle j|)\parallel\,\le\,\parallel(\mathcal{E}_{T}-e^{\tilde{\mathcal{K}}_{eff}})\mathcal{P}_{0}\parallel=O(1/T)$,
showing that effectively, one can generate unitary gates on the logical
qubit space ${\cal C}$ up to an error $1/T.$ In view of Remark \textbf{{3)}}
one is allowed to add to $\mathcal{L}^{\alpha}$ any perturbation
$\mathcal{V}$ satisfying $\mathcal{P}_{0}\mathcal{V}\mathcal{P}_{0}=0$,
and still obtain the same unitary gates $u^{\alpha}$ within an error
$c/T$ albeit with a possibly different $c$ \cite{inpreparation}.  In \cite{SM} we show
the stability of this dynamics also against certain {\em dissipative}
perturbations of ${\cal L}_{0}.$
Fig.~\ref{fig:J_zero} (left panel)
shows that the whole $14$-dimensional SSM is evolving unitarily in
the long time limit. 

To illustrate our results let us consider the experimental DFS system studied in \cite{DFS-exp} consisting of a couple of trapped $^9Be^+$ ions subject to collective dephasing
[$\gamma_{x,y}=0$ in (\ref{L0-four})].  In this case  $\tau_R\sim 5\,\mu s$ and (assuming a similar relaxation time for a four qubits system)  Eq.~(\ref{op-time}) and Fig.~\ref{fig:J_zero} show that for $T\sim 500 \,\mu s$ one should observe small deviations of the effective dynamics from unitarity.

\prlsection{Unitaries over noiseless subsystem} Next we discuss dissipation-assisted computation over noiseless subsystems \cite{NS}. 
The Liouvillian is in the class previously discussed, $\mathcal{L}_{0}\left(\rho\right)=\Phi\left(\rho\right)-\rho$,
taking 
$\Phi\left(\rho\right)=\frac{1}{3}\sum_{\alpha=1}^{3}U_{\alpha}\rho U_{\alpha}^{\dagger},\quad U_{\alpha}=e^{i\phi_{\alpha}S^{\alpha}},$
where $S^{\alpha}$ are again collective spin operators. 
For generic $\phi_{\alpha}$'s the SSM coincides with one of the former
examples i.e, rotationally invariant state. The latter for an odd
number $N$ of spins, contains only mixed states. 
  As perturbation we use the following Hamiltonian $H=\sigma_{1}^{x}\sigma_{2}^{x}$
and the full Liouvillian reads $\mathcal{L}=\mathcal{L}_{0}-i\vartheta/T\left[H,\bullet\right]$.
Again one observes an effective unitary evolution, up to an error
$O(1/T),$ (see Fig.~\ref{fig:J_zero} right panel) over the full five-dimensional
SSM; in particular, 
this construction can be seen as a scheme to enact dissipation-assisted
control over the noiseless-subsystem ${\mathbf{C}}^{2}$ factor
\cite{ogy}.

In \cite{NS-exp} noiseless-subsystems  have been realized in a NMR system comprising three nuclear spins subject to collective (artificial) noise;
for a relaxation time $\tau_R < 1/30\,s $ the noiseless encoding provides an advantage. Fig~\ref{fig:J_zero}  shows that setting the operation time, say at
$T=100\,\tau_R,$ then effective dynamics over the NSs becomes very close to a unitary one.

Finally we would like to stress that the Markovian form (1) is just  {sufficient}  (and mathematically convenient) to prove the existence of an effective projected dynamics, {\em{not}} necessary.  The spin-boson Hamiltonian  discussed in \cite{SM} 
indicates that the relevant dynamical mechanism is the  existence of a {\em{strong}} system-bath coupling that adiabatically-decouples non steady-states from the dynamics.

\prlsection{Conclusions}\label{conclusions} 
In this Letter we have shown how an effective unitary dynamics can be enacted over the manifold of steady states  of a strongly-dissipative system. The strategy is to introduce a small time-rescaled Hamiltonian 
term in the system's Liouvillian largely dominated by the dissipative
processes. In the long time limit the dynamics leaves the steady state
manifold invariant and becomes unitary up to a small error whose strength
is connected to the Liouvillian relaxation time, and total operation time. The effective Hamiltonian ruling the
long time-dynamics is shaped by the continuous interplay of the weak
Hamiltonian control with the the fast relaxation process that adiabatically
decouples non-steady states. This effective projected Hamiltonian,
in some cases, can be seen as a symmetrized form of the bare one and
it is robust against all perturbations, dissipative or Hamiltonian,
that are filtered out by this environment-induced symmetrization.

To illustrate these ideas we have shown how to realize quantum gates
on \mbox{steady-state} manifolds comprising decoherence-free subspaces
\cite{DFS} as well as noiseless subsystems \cite{NS}. In all these
cases we have also provided a numerical estimate of the deviations
from the ideal long-time unitary behavior and the actual, finite time,
one. Agreement with the theoretical prediction (\ref{projection-th})
is found in all cases. 

The results of this Letter seem to suggest the
intriguing possibility of fighting quantum decoherence by introducing 
even more quantum decoherence.

\begin{acknowledgments}
This work was partially supported by the ARO MURI grant W911NF-11-
1-0268 and by NSF grant PHY- 969969. Useful input from S. Garnerone,
J. Kaniewski, D. Lidar, I. Marvian and S. Muthukrishnan is gratefully
acknowledged.
\end{acknowledgments}

\clearpage
\appendix





\section{Proof of Main Theorem}

\label{lorenz-proof} 

In this section we provide a proof of Eq.~(2) of the main text. Our
approach and terminology rely heavily on the classical text \cite{kato}.
Let ${\cal L}_{T}={\cal L}_{0}+T^{-1}\tilde{{\cal K}}$. For $T^{-1}=0$,
${\cal L}_{0}$ is assumed to have a degenerate steady state manifold,
i.e.~$\dim\ker\mathcal{L}_{0}=m_{0}>1.$ For small non-zero $T^{-1}$,
some eigenvalues of $\mathcal{L}_{T}$ (may) depart from $\lambda=0$.
The set of these eigenvalues is called the $\lambda$-group since
they cluster around the unperturbed eigenvalue, in this case $\lambda=0$,
for small $\left|T^{-1}\right|$ \cite{kato}. Let ${\cal P}$ be
the projection associated to the $\lambda$-group originating from
the degenerate $\lambda=0$ eigenvalue of ${\cal L}_{0}$ (whose associate
projection is given by ${\cal P}_{0}$). Define also the projected
Liouvillian ${\cal L}_{r}:={\cal P}{\cal L}_{T}{\cal P}$. A central
result of \cite{kato} states that both $\mathcal{P}$ and $\mathcal{L}_{r}$
are analytic in $T^{-1}$, i.e.~their power series in $T^{-1}$ have
a finite radius of convergence. Since $\mathcal{P}$ commutes with
$\mathcal{L}_{T}$ one clearly has $e^{t{\cal L}_{T}}{\cal P}=e^{t{\cal L}_{r}}{\cal P}$.
We can now expand both $\mathcal{P}$ and $\mathcal{L}_{r}$ around
$T^{-1}=0$. Accordingly we write 
\begin{equation}
e^{t{\cal L}_{T}}({\cal P}_{0}+\delta{\cal P})=e^{t\,{\cal L}_{\mathrm{eff}}+t\,\delta{\cal L}}({\cal P}_{0}+\delta{\cal P}),
\end{equation}
where we defined $\delta{\cal P}:={\cal P}-{\cal P}_{0},\,{\cal L}_{\mathrm{eff}}:={\cal P}_{0}{\cal L}_{T}{\cal P}_{0}=T^{-1}{\cal P}_{0}\tilde{{\cal K}}{\cal P}_{0},$
and $\delta{\cal L}:={\cal L}_{r}-{\cal L}_{\mathrm{eff}}.$ Using
\cite{kato} and remembering that $t\le T$, one finds 
\begin{equation}
\|\delta{\cal P}\|=O(1/T),\;\;\|t\,\delta{\cal L}\|=O(1/T),\;\;\|t\,{\cal L}_{\mathrm{eff}}\|=O(1).\label{norms}
\end{equation}

For example, in case the zero eigenvalue has no nilpotent part, as
it happens in physical systems, one has \cite{kato} 
\begin{align}
{\cal L}_{r} & ={\cal P}_{0}{\cal L}_{T}{\cal P}_{0}\nonumber \\
 & -T^{-2}\Big({\cal P}_{0}\tilde{{\cal K}}{\cal P}_{0}\tilde{{\cal K}}{\cal S}+{\cal P}_{0}\tilde{{\cal K}}{\cal S}\tilde{{\cal K}}{\cal P}_{0}\nonumber \\
 & +{\cal S}\tilde{{\cal K}}{\cal P}_{0}\tilde{{\cal K}}{\cal P}_{0}\Big)+O(T^{-3})\label{deg-pert-th}
\end{align}
and 
\begin{equation}
\delta\mathcal{P}=-T^{-1}\left({\cal P}_{0}\tilde{{\cal K}}{\cal S}+{\cal S}\tilde{{\cal K}}{\cal P}_{0}\right)+O(T^{-2}).\label{eq:dP}
\end{equation}
In Eqns.~(\ref{deg-pert-th}) and (\ref{eq:dP}) above, ${\cal S}$
is the projected resolvent of\emph{ $\mathcal{L}_{0}$} related to
the $\lambda=0$ eigenvalue. Explicitly, if $\mathcal{L}_{0}$ has
the following Jordan decomposition 
\begin{equation}
\mathcal{L}_{0}=\sum_{j=0}^{s-1}\lambda_{j}\mathcal{P}_{j}+\mathcal{D}_{j},
\end{equation}
with $\mathcal{P}_{j}$ projectors, $\mathcal{D}_{j}$ nilpotents
and $\lambda_{0}=0$, the projected resolvent is given by 
\[
\mathcal{S}=-\sum_{j=1}^{s-1}\left[(-\lambda_{j})^{-1}\mathcal{P}_{j}+\sum_{n=1}^{m_{j}-1}(-\lambda_{j})^{-n-1}\mathcal{D}_{j}^{n}\right].
\]

Define further $\Delta:=e^{t\,{\cal L}_{\mathrm{eff}}+t\,\delta{\cal L}}-e^{t\,{\cal L}_{\mathrm{eff}}}$.
Now we use the inequality $\|e^{X+Y}-e^{X}\|\le\|Y\|e^{\|X\|+\|Y\|}$
with $X=t\,{\cal L}_{\mathrm{eff}}$ and $Y=t\,\delta{\cal L},$ to
obtain 
\begin{equation}
\|\Delta\|\le\|t\,\delta{\cal L}\|e^{\|t\,{\cal L}_{\mathrm{eff}}\|+\|t\,\delta{\cal L}\|}.\label{eq:bound}
\end{equation}
Therefore 
\begin{eqnarray}
e^{t\,{\cal L}}{\cal P}_{0} & = & e^{t\,{\cal L}_{\mathrm{eff}}}{\cal P}_{0}+{\cal E},\\
\mathrm{with}\quad\mathcal{E} & = & \Delta(\mathcal{P}_{0}+\delta{\cal P})+(e^{t\,{\cal L}_{\mathrm{eff}}}-e^{t\,{\cal L}})\delta{\cal P}.
\end{eqnarray}
The proof is completed using triangle inequality and the bounds (\ref{norms})
and (\ref{eq:bound}) (and setting $t=T$), implying $\|{\cal E}\|=O(1/T)$.

Note that this proof, together with the bound (2) in the main text,
remains valid in a slightly more general setting where the eigenvalues
$\lambda_{h}$ of $\mathcal{L}$ satisfy $\mathrm{Re}(\lambda_{h})\le0$.
For example, in the extreme case of unitary dynamics where the eigenvalues
are purely imaginary this result become essentially the standard adiabatic
theorem as discussed in Sec.~\ref{sec:Hamiltonian-example}, but
intermediate cases are accounted for as well.

We now consider a case in which ${\cal P}_{0}\tilde{{\cal K}}{\cal P}_{0}=0$.
Performing the rescaling $\mathcal{K}=T^{1/2}\tilde{\mathcal{K}}$
one is led to analyze $\mathcal{L}_{r}={\cal L}_{\mathrm{eff}}+\delta\mathcal{L}$
with ${\cal L}_{\mathrm{eff}}=-T^{-1}{\cal P}_{0}\tilde{{\cal K}}{\cal S}\tilde{{\cal K}}{\cal P}_{0}$.
The bounds in Eq.~(\ref{norms}) become now 
\begin{equation}
\|\delta{\cal P}\|=O(1/T^{1/2}),\;\;\|t\,\delta{\cal L}\|=O(1/T^{1/2}),\;\;\|t\,{\cal L}_{\mathrm{eff}}\|=O(1).\label{norms-1}
\end{equation}
Reasoning as previously we now obtain $e^{T\,{\cal L}}{\cal P}_{0}=e^{T\,{\cal {\cal L}_{\mathrm{eff}}}}{\cal P}_{0}+{\cal E}$
with $\|{\cal E}\|=O\left(T^{-1/2}\right)$.


\section{Hamiltonian example\label{sec:Hamiltonian-example}}

\begin{figure}
\noindent \begin{centering}
\includegraphics{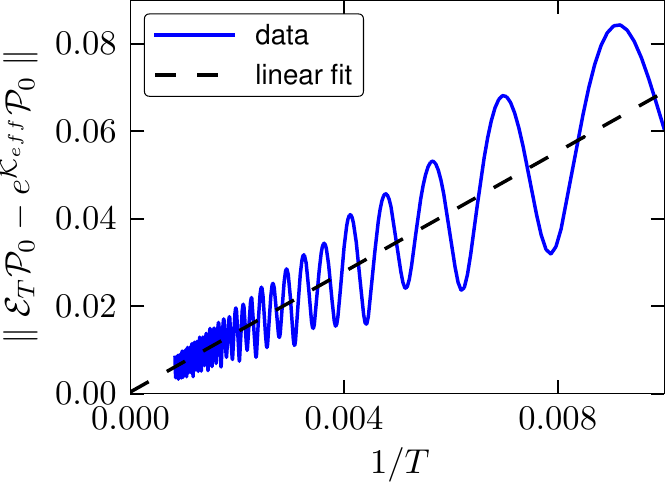} 
\par\end{centering}

\protect\caption{Distance of the effective unitary evolution from the exact (unitary)
one one as a function of $1/T$. The norm used is the maximum singular
value of the maps realized as matrices over ${\cal H}^{\otimes\,2}.$
Parameter used are $\omega_{k}=k$ ($k=2\pi n/N_{B}$, $n=1,\ldots,N_{B}$),
$g_{k}=0.045$ and $\vartheta=1$. The system has $N_{S}=3$ and $N_{B}=60$.
\label{fig:Distance-multi}}
\end{figure}

as reminded in the previous section, our projection result Eq.~(2)
of the main text, holds also when $\mathcal{L}_{0}=-i\left[H_{0},\bullet\right]$
and in this case it simply amounts to a type of adiabatic theorem
for closed quantum systems. To illustrate this fact we consider a
system of $N_{S}$ spins interacting collectively with $N_{B}$ bosons
and Hamiltonian 
of atoms with the radiation field. We restrict ourself to the space
of only one boson or spin excitation Hilbert space\emph{ $\mathcal{H}=\mathrm{span}\left\{ |x\rangle_{S}|0\rangle_{B},|\Downarrow\rangle_{S}|k\rangle_{B}\right\} $},
$x=1,\ldots,N_{s},$ $k=1,\ldots,N_{b}$ where $|x\rangle_{S}|0\rangle_{B}=S_{x}^{+}|\Downarrow\rangle_{S}|k\rangle_{B}$,
$|\Downarrow\rangle_{S}$ has all spins down, $|k\rangle_{B}$ one
boson in mode $k$ and $|0\rangle_{B}$ is the boson vacuum. 
Hamiltonian $H_{0}$ admits the following dark states ($H_{0}|\psi_{q}\rangle=0$),
$|\psi_{q}\rangle=N_{S}^{-1/2}\sum_{x=1}^{N_{S}}e^{-i2\pi qx/N_{s}}|x\rangle_{S}|0\rangle_{B}$,
with $q=1,\ldots,N_{s}-1$ \cite{zan97}. 
SSM includes all the states built over the dark state manifold all
of which are decoherence-free at zero temperature \cite{zan97}. Let
us now introduce an Hamiltonian perturbation which conserves the total
number of excitations, such as $H_{1}=\sigma_{1}^{z}$, and the corresponding
superoperator $\mathcal{K}=-i\vartheta/T\left[H_{1},\bullet\right]$.
projected Hamiltonian over the dark-state manifold turns out to be
$\tilde{K}_{eff}=\vartheta[2(N_{S}-1)/N_{s}|\phi\rangle\langle\phi|-\openone]$
with $|\phi\rangle=(N_{S}-1)^{-1/2}\sum_{k=1}^{N_{s}-1}e^{-i2\pi k/N_{S}}|\psi_{k}\rangle$,
shows how Eq.~(2) of the main text is fulfilled in this unitary case
as well.



\begin{thebibliography}{10}
\bibitem{norm} In this paper, unless otherwise stated the norms for
superoperators ${\cal M}$ will be 
$\|{\cal M}\|:=\sup_{\|X\|_{1}=1}{\|{\cal M}(X)\|_{1}}.$ 
For semigroups $\{{\cal E}_{t}\}_{t\ge0}$ of CP-maps one has $\|{\cal E}_{t}\|\le1.$
$\|X\|$ will denote the standard operator norm for $X\in$L$({\cal H}).$
\bibitem{commutant}
By definition ${\cal A}^\prime=:\{X\,/\, [X,O]=0,\,\forall O\in{\cal A}\}.$
 Standard structure-theorems for $C^{*}$-algebras
imply that 
${\cal A}^{\prime}\cong\bigoplus_{J}{\mathrm{L}}({\mathbf{C}}^{n_{J}})\otimes\openone_{d_{J}}$
 where $J$ labels the irreducible representations of ${\cal A}$
with dimension $d_{J}$ and multiplicity $n_{J}$ \cite{NS}. 
In this case ${\cal P}_{0}(X)=\int dU\, UXU^{\dagger}$
where the Haar-measure integral is performed over the unitary group
of the algebra ${\cal A}.$
\bibitem{proj-comm} If $X={\cal P}_{0}(X)$ one has : ${\cal P}_{0}\,\tilde{{\cal K}}\,{\cal P}_{0}(X)=-i{\cal P}_{0}([\tilde{K},X])=-i{\cal P}_{0}(\tilde{K}X)+i{\cal P}_{0}(X\tilde{K})=-i{\cal P}_{0}(\tilde{K})X+iX{\cal P}_{0}(\tilde{K})=-i[{\cal P}_{0}(\tilde{K}),\, X].$
Where we used e.g., ${\cal P}_{0}(\tilde{K}X)={\cal P}_{0}(\tilde{K})X$
valid as $X$ is  in ${\cal A}^{\prime}$ \cite{commutant}.
\bibitem{simple-calcu} Let $X\in{\mathrm{L}}({\cal C})$ and consider
${\cal P}_{0}\,\tilde{{\cal K}}\,{\cal P}_{0}(X)=-i{\cal P}_{0}([\tilde{K},\, X]).$
One can write $\tilde{K}=\Pi\tilde{K}+Q\tilde{K}$ where $Q:=1-\Pi.$
Therefore ${\cal P}_{0}(\tilde{K}X)={\cal P}_{0}(\Pi\tilde{K}X+Q\tilde{K}X)=\Pi\tilde{K}X=(\Pi\tilde{K}\Pi)X$
where we have used the properties of ${\cal P}_{0}$ assumed in the
main text and $X=\Pi X.$ Considering now the $X\tilde{K}$ term in
the commutator in the same way one finds ${\cal P}_{0}([\tilde{K},\, X])=[\Pi\tilde{K}\Pi,\, X]$
for all $X\in{\mathrm{L}}({\cal C}).$
\end{thebibliography}
\end{document}